\documentclass[journal,comsoc]{IEEEtran}
\usepackage{graphicx}
\usepackage{tabularx}
\usepackage{caption}
\usepackage{epstopdf}
\usepackage{float}
\usepackage{bm,upgreek}
\usepackage{amsfonts}
\usepackage{algorithm,algpseudocode}
\usepackage{enumitem}
\usepackage{mathtools}
\usepackage[caption=false]{subfig}
\usepackage{multirow}
\usepackage{booktabs}
\usepackage[subnum]{cases}
\usepackage{setspace}

\usepackage{algorithm,algpseudocode}
\usepackage{cite}
\usepackage{makecell}
\usepackage{color}
\usepackage{lipsum}
\usepackage{amsmath}
\usepackage{setspace}
\usepackage{amssymb}
\usepackage{acronym}
\usepackage{url}
\usepackage[labelfont=bf]{caption}
\usepackage[T1]{fontenc}
\usepackage{amsmath}
\usepackage[cmintegrals]{newtxmath}
\usepackage{hyperref}

\begin{document}

\title{Visible Light Communications Based Indoor Positioning via Compressed Sensing}

\author{Kristina Gligori{\'c}, Manisha Ajmani, Dejan Vukobratovi{\'c} and Sinan Sinanovi{\'c}
\thanks{Kristina Gligori{\'c} and Dejan Vukobratovi{\'c} are with Faculty of Technical Sciences, University of Novi Sad, Serbia, (email: kristinagligoric.ns@gmail.com, dejanv@uns.ac.rs).}
\thanks{Manisha~Ajmani and Sinan~Sinanovi{\'c} are with the School of Engineering and Built Environment, Glasgow Caledonian University, Glasgow, G4 0BA, UK, (e-mail: manisha.ajmani@gcu.ac.uk, sinan.sinanovic@gcu.ac.uk).}
\thanks{S. Sinanovi{\'c} acknowledges support from Digital Health $\&$ Care Institute. D. Vukobratovi{\'c} received funding from EU H2020 research and innovation programme under the Marie Sklodowska-Curie grant agreement No 734331.}
}
\maketitle

\begin{abstract}
This paper presents an approach for visible light communication-based indoor positioning using compressed sensing. We consider a large number of light emitting diodes (LEDs) simultaneously transmitting their positional information and a user device equipped with a photo-diode. By casting the LED signal separation problem into an equivalent compressed sensing framework, the user device is able to detect the set of nearby LEDs using sparse signal recovery algorithms. From this set, and using proximity method, position estimation is proposed based on the concept that if signal separation is possible, then overlapping light beam regions lead to decrease in positioning error due to increase in the number of reference points. The proposed method is evaluated in a LED-illuminated large-scale indoor open-plan office space scenario. The positioning accuracy is compared against the positioning error lower bound of the proximity method, for various system parameters.
\end{abstract}


%
\IEEEpeerreviewmaketitle
\section{Introduction}
\IEEEPARstart{I}{ndoor} tracking and positioning systems have been gaining traction among the research community after the huge success of outdoor tracking and positioning systems based on global positioning system (GPS) technology \cite{up1}. However, GPS is not suitable for indoor environments owing to its incapability to penetrate walls of a building. Other indoor positioning systems (IPS) based on ultrasound, radio frequency identification (RFID), and wireless local area network (WLAN) require additional resources for their operation, thus increasing the overall system cost \cite{paper7,up2}.  

Visible light communication (VLC) with light emitting diodes (LEDs) is a relatively new technology that became the subject of exploration within academic and industrial research in the field of indoor positioning \cite{paper8},\cite{VLCpossur}. Its advantages include high bandwidth, high positioning accuracy, license free operation and immunity to electromagnetic interference \cite{pap6}. It can be integrated with the existing lighting infrastructure for its operation. VLC can be used for various applications ranging from navigation and emergency rescue to object localization. 

There are two major components of the VLC-based IPS: LED transmitters and photo-diode (PD) optical receivers. LED transmitters are installed at fixed locations inside a room and continuously transmit their position coordinates encoded as an optical signal. When a mobile optical receiver comes in the vicinity of a LED transmitter, it receives the optical signal and decodes the LED position to estimate its own position. The transmitted optical signals are affected by multipath propagation, shadowing, and interference from various noise sources \cite{paper9}. Thus, considerable amount of research has been done to find suitable implementation techniques for the VLC-based IPS which helps in achieving high positioning accuracy.

Most of the research on VLC-based IPS investigates techniques based on triangulation/trilateration, fingerprinting or proximity methods \cite{VLCpossur}. These approaches usually use features extracted from signal such as received signal strength (RSS), angle of arrival (AOA)\cite{s22}, or time of arrival (TOA) \cite{paper10}. In most of the methods, authors assume that power from two or more different light sources can be easily separated. Also, to remove angular dependency in RSS based localization, some studies assume fixed height of the receiver and the receiver surface being parallel to the transmitter \cite{s12}. In practice, such assumptions are questionable as a mobile receiver cannot be constrained by a fixed angle or height. Additional problems such as system complexity, computational and implementation costs makes it difficult to implement these techniques.

When the light beam regions of multiple LEDs in the IPS intersect, given the underlying optical wireless communication channel model, the problem of estimating signal strength can be translated to finding the channel gains of the set of received signals. Considering that most of the channel gains at the optical receiver will be zero, as the receiver can receive signal from only a limited number of nearby LEDs, the resulting problem reduces to an equivalent compressed sensing (CS) problem and can be efficiently solved using sparse signal recovery methods \cite{CSbook}. In this letter, the feasibility of using CS for signal separation is explored and a new indoor position estimation approach based on proximity method is devised following the intuition that overlapping light beam regions improve the accuracy of VLC based IPS due to increased number of reference positions. The proposed CS-based method is investigated in a realistic LED-illuminated large-scale indoor open-plan office space environment. For various system parameters, the position accuracy of the proposed method is evaluated and compared against the performance limits achievable using the proximity method. 

This letter is organized as follows: Section \ref{sec:2} discusses the design of IPS using multiple LEDs. Section \ref{sec:3} discusses the proposed CS-based localization technique in detail followed by the performance analysis and numerical results in Section \ref{sec:4}. The letter ends with concluding remarks. 

\section{Indoor Positioning via Multiple LEDs}
\label{sec:2}

\begin{figure}[htbp]
\vspace{-10pt}
\center
  \includegraphics[width=0.47\textwidth]{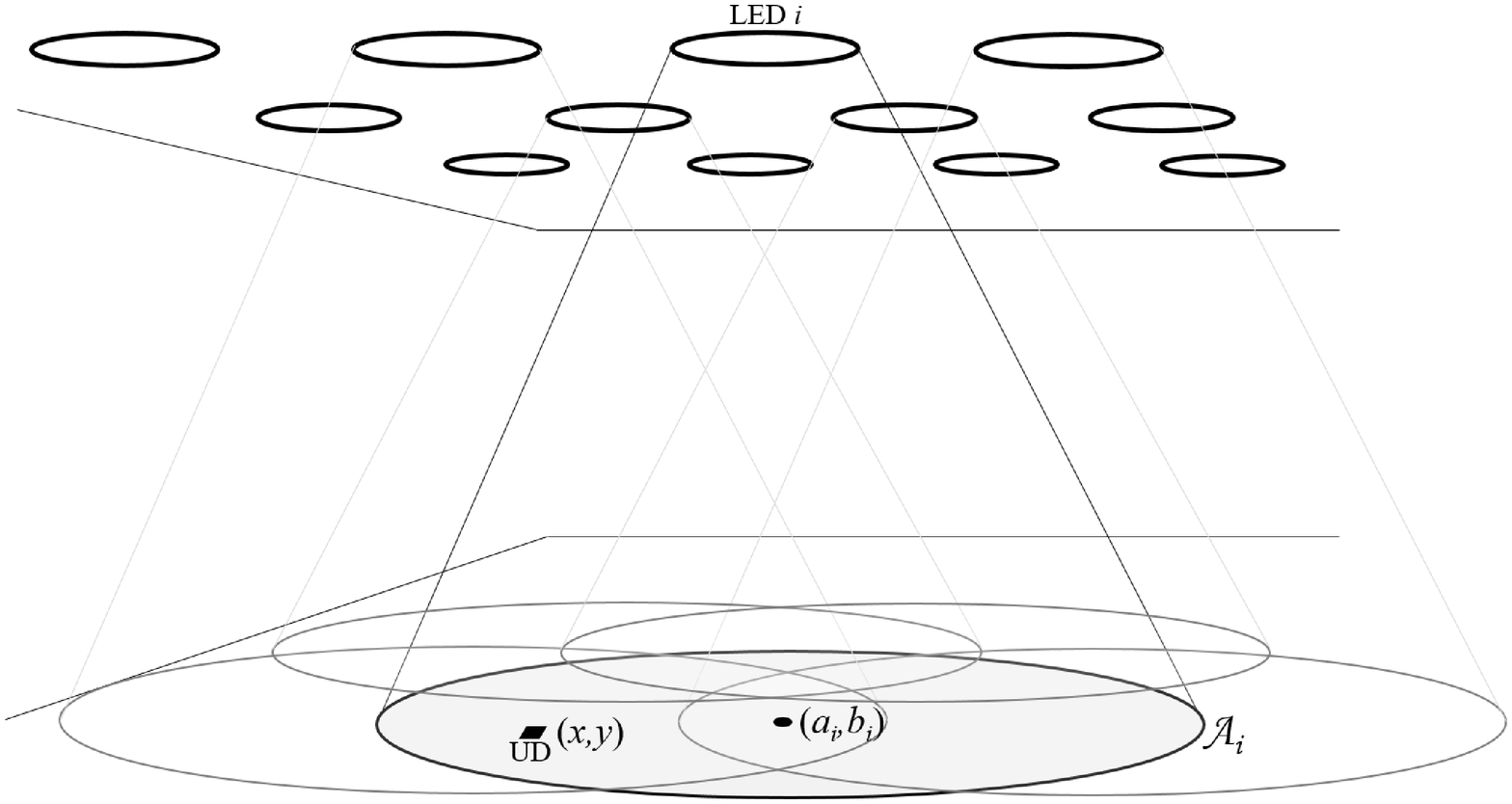}
\caption{Multiple-LED position estimation model.}
\label{Fig:1} 
\vspace{-7pt}
\end{figure}

For the purpose of analysis, a large indoor LED-illuminated environment (e.g., an open-space office or an industrial hall) with $N$ LED sources attached to the ceiling is considered. LEDs are placed on a grid of known positions $\mathbf{z}_i=[a_i,b_i,h]$, where $a_i$ and $b_i$ are the $i$-th LED coordinates of the floor/ceiling plane, while $h$ is the height of all LEDs from the ground. It is assumed that the $i$-th LED illuminates a certain coverage area $\mathcal{A}_i$ of the floor plane $\mathcal{F}$, as shown in Fig. \ref{Fig:1}. A user device (UD) equipped with a PD is placed at an unknown location $\mathbf{u}=[x,y,0],$ (for simplicity, we assume $\mathbf{u} \in \mathcal{F}$). Focusing on the line-of-sight (LOS) communication and ignoring the signal reflections, we assume the PD is able to detect the signal from the $i$-th LED if $\mathbf{u} \in \mathcal{A}_i$. In general, the coverage areas of different LEDs overlap, thus leading to a positioning problem via multiple-LED estimation model (MLEM) recently considered in \cite{ola}. 

For the purpose of positioning, we propose that, at a given time instant (e.g., periodically or upon request), all LEDs simultaneously transmit their binary ID sequences $\mathbf{s}_i=(s_{i,1},s_{i,2},\ldots,s_{i,M})^T$, $1 \leq i \leq N,$ of length $M$ bits. Each bit of the ID sequence $\mathbf{s}_i$ is generated independently and uniformly at random and we assume that the UD knows the ID sequences of all LEDs. The LED signal is transmitted using a simple on-off keying (OOK) modulation with the symbol period selected sufficiently large so that we can: i) assume all LED transmissions are perfectly synchronized, and ii) ignore all distance-dependent delays. At an unknown position, the received signal $\mathbf{y}=(y_{1},y_{2},\ldots,y_{M})^T$ at the UD represents a superposition of signals arriving from surrounding LEDs:
\begin{equation}
\label{eq1}
\mathbf{y} = \sum_{i=1}^{N} \lambda_i \alpha_i \mathbf{s}_i + \mathbf{n},
\end{equation}
where $\alpha_i$ is the gain of the optical channel\footnote{For the optical LOS channel gain, we adopt the following model: 
\begin{equation}
\label{eq2}
       \alpha_\textit{i}  = 
        \begin{cases}
            \frac{A_r(m+1)}{2\pi{d_{i}^2}}\mathrm{cos}^m(\phi_{i}){g}(\psi_{i}){T_s}(\psi_{i})\mathrm{cos}(\psi_{i}), & \text{ ${0}\leq{\psi_{i}}\leq{\psi_c}$ } \\
            0, & \text{otherwise}
        \end{cases}
\end{equation}
\begin{equation} 
\label{eq3}
m = \frac{-\ln{2}}{\mathrm{\ln{cos}}(\Phi_{1/2})}, \\ ~~~
\mathrm{g}(\psi)  = 
        \begin{cases}
            \frac{n_{r}^2}{\mathrm{sin^2}(\psi_c)}, & \text{for ${0}\leq{\psi_{i}}\leq{\psi_c}$ $$} \\
            0, & {0}\geq{\psi_c}
        \end{cases}
\end{equation}
where $A_r$ is physical area of the detector, $m$ is Lambertian LED order, ${T_{s}}(\psi)$ is the gain of the optical filter, ${g}(\psi)$ is the gain of the optical concentrator, $\psi_c$ denotes the field of vision (FOV) of the receiver, $n_{r}$ is the refractive index of the concentrator, $\Phi_{1/2}$ is semi-angle at half power of an LED, $d_i$, $\phi_i$ and $\psi_i$ are the distance, radiation and incidence angle between the $i$-th LED and the PD, respectively \cite{OWCbook}.} between the $i$-th LED and the UD (assumed to remain constant during $M$ symbol intervals), $\mathbf{n}=(n_{1},n_{2},\ldots,n_{M})^T$ is a vector of additive i.i.d. Gaussian noise samples, and $\lambda_i=\mathbb{I}_C (\mathbf{u} \in \mathcal{A}_i)$ is a coverage indicator variable equal to one if the UD is covered by the $i$-th LED. Representing \eqref{eq1} in the matrix form, we get:
\begin{equation}\label{eq4}
\mathbf{y} = \mathbf{S}\mathbf{x} + \mathbf{n},
\end{equation}
where $\mathbf{S}=(\mathbf{s}_1,\mathbf{s}_2,\ldots,\mathbf{s}_N)$ is an $M \times N$ binary matrix containing all LED ID sequences organized into columns. In addition, we compactly define: 
\begin{equation}
\mathbf{x} = \mathbf{\lambda}\odot \mathbf{\alpha} \triangleq (\lambda_1\alpha_1,\lambda_2\alpha_2,\ldots,\lambda_N\alpha_N)^T,
\end{equation}
where $\mathbf{\lambda}=(\lambda_1,\lambda_2,\ldots,\lambda_N)^T$ is the coverage indicator vector and $\mathbf{\alpha}=(\alpha_1,\alpha_2,\ldots,\alpha_N)^T$ is the optical channel gain vector.

The problem we consider is that of estimating the unknown coordinates $\mathbf{u}$ of the UD based on the signal $\mathbf{y}$ received from LED transmitters and knowledge of the matrix $\mathbf{S}$ of LED ID signatures. We split this problem into two steps: i) \emph{LED ID detection and channel estimation} step, where we recover the vector $\mathbf{x}$ from \eqref{eq4} using ideas from CS, and ii) \emph{UD position estimation} step, where, based on the recovered vector $\mathbf{x}$, we estimate the UD position $\mathbf{u}$.    

\section{VLC-based Positioning via Compressed Sensing}
\label{sec:3}
\subsection{LED ID Detection and Channel Estimation}

In a large indoor environment, we assume that any LED coverage area $\mathcal{A}_i$ illuminates only a small fraction of the floor plane $\mathcal{F}$. Thus, at any location $\mathbf{u}$, the UD will be covered by a small number $K=K(\mathbf{u})$ of surrounding LEDs, where $K=|\mathcal{S}_{\mathbf{\lambda}}|$ is the cardinality of the support set $\mathcal{S}_{\mathbf{\lambda}}$ (i.e., the set of indices of non-zero elements) of the coverage vector $\mathbf{\lambda}$. Consequently, the vector $\mathbf{x}$ is a $K$-sparse vector, meaning that it contains $K$ non-zero components. Assuming $K \ll N$, equation \eqref{eq4} represents a canonical noisy CS problem \cite{CSbook}. The theory of CS provides conditions under which the vector $\mathbf{x}$ can be recovered at the UD from the received signal $\mathbf{y}$ and knowledge of $\mathbf{S}$. In general, the $K$-sparse vector $\mathbf{x}$ of length $N$ can be recovered from the vector $\mathbf{y}$ of length $M=O(K\log{\frac{N}{K}}) \ll N$, if the matrix $\mathbf{S}$ satisfies certain properties (e.g., restricted isometry property), which are known to hold for random Bernoulli matrices \cite{CSbook}.

From $\mathbf{y}$ and $\mathbf{S}$, the UD applies suitable CS recovery method to generate an estimate $\hat{\mathbf{x}}$ of the vector $\mathbf{x}$. From $\hat{\mathbf{x}}$, the UD estimates the support set $\hat{\mathcal{S}}_{\mathbf{\lambda}}$ of $\mathbf{x}$, which completes the LED ID detection part. The components of $\hat{\mathbf{x}}$ that correspond to the estimated support set $\hat{\mathcal{S}}_{\mathbf{\lambda}}$  provide estimated channel gains: $\hat{\mathbf{\alpha}}_{\hat{\mathcal{S}}_{\mathbf{\lambda}}} = \{\hat{x}_i, i \in \hat{\mathcal{S}}_{\mathbf{\lambda}}\}.$ We also note \emph{the group-sparsity property} of the LED ID detection problem: the set of non-zero coordinates of $\mathbf{x}$ is highly spatially correlated. 

For the range of the number of LEDs $N$ in our setup, classical methods such as basis pursuit denoising (BPDN), or its low-complexity alternatives such as orthogonal matching pursuit (OMP), can be applied \cite{CSbook}.
In this paper, we apply OMP as it is well-understood and investigated. OMP recovers $\mathbf{x}$ via fast, greedy iterative procedure, by identifying sequentially the columns of $\mathbf{S}$ that are most strongly correlated with the remaining part of $\mathbf{y}$. Proper sequence of column indices minimizes the residual which eventually leads to the convergence of OMP, as detailed in \cite{comp2}.

\subsection{UD Position Estimation}

After recovering vector $\mathbf{x}$ using CS, the next step is to estimate the UD position. Due to complexity of taking the unknown angles into account, we resort to a simplified \emph{proximity method} that uses only the estimated support set $\hat{\mathcal{S}}_{\mathbf{\lambda}}$ for position estimation, thus effectively disregarding the channel gains \cite{proximity}. The estimated UD position $\hat{\mathbf{u}}=[\hat{x},\hat{y}]$ is obtained as the centroid of an area $\mathcal{A}_c$ obtained as a union of LED coverage areas of all detected LEDs, i.e., $\mathcal{A}_c = \bigcup_{i \in \hat{\mathcal{S}}_{\mathbf{\lambda}}} \mathcal{A}_i$. With the sufficient LED density and overlapping of LED beams, this low-cost method provides high positioning accuracy in indoor environment, as detailed next.

\subsection{VLC-based positioning using CS: The proposed algorithm}

The proposed low-complexity CS-based VLC positioning algorithm is presented below. Besides the received signal $\mathbf{y}$ and the matrix $\mathbf{S}$, the UD requires knowledge of system geometry: LED locations and coverage areas $\{\mathbf{z}_i, \mathcal{A}_i\}_{1\leq i \leq N}$, maximum signal sparsity $K_{max}$ at any point of the floor plane: $K_{max}=\max_{\mathbf{u} \in \mathcal{F}} K(\mathbf{u})$, and suitably selected distance threshold $d_{th}$.

\begin{algorithm} [ht]
\caption{VLC-based Positioning via CS}
\label{VLC-Pos}
\begin{spacing}{1.25}
\begin{algorithmic}[1] 

\Procedure {Noisy OMP Recovery}{}
\State Inputs: $\mathbf{y}$, $\mathbf{S}$;
\State Estimate  $\hat{\mathbf{x}}$ = OMP($\mathbf{y}$, $\mathbf{S}$) (see \cite{comp2} for details);
\State Output: $\hat{\mathbf{x}}$;
\EndProcedure

\Procedure {UD Position Recovery}{}
\State Inputs: $\hat{\mathbf{x}}$, $\{\mathbf{z}_i, \mathcal{A}_i\}$, $K_{max}$, $d_{th}$;
\State $\hat{\mathbf{x}}_{s}$, $\mathbf{i}_{s}$ = SORT($\hat{\mathbf{x}}$); (sorts $\hat{\mathbf{x}}$ descending into $\hat{\mathbf{x}}_{s}$, and $\mathbf{i}_{s}$ are original indices after sorting) 
\State Initialize $\hat{\mathcal{S}}_{\mathbf{\lambda}} = \{\mathbf{i}_s(1)\}$; ($\mathbf{i}_s(1)$ is the index of the strongest detected LED)
\State Initialize $\hat{\mathbf{u}}=[\hat{x}, \hat{y}] = \mathbf{z}_{\mathbf{i}_s(1)}$;
 
\For{j = 2 : $K_{max}$}
\If{$|\mathbf{z}_{\mathbf{i}_s(j)} - \hat{\mathbf{u}}|_{2} < d_{th}$}
\State $\hat{\mathcal{S}}_{\mathbf{\lambda}} = \{\hat{\mathcal{S}}_{\mathbf{\lambda}}, \mathbf{i}_s(j)\}$;
\State $\hat{\mathbf{u}}=[\hat{x}, \hat{y}]$ = PROX$\left(\{\mathbf{z}_i\}_{i \in \hat{\mathcal{S}}_{\mathbf{\lambda}}}\right)$;
\EndIf
\EndFor
\State Output: $\hat{\mathcal{S}}_{\mathbf{\lambda}}$; (Set of LED IDs covering UD)
\State Output: $\hat{\mathbf{u}}$; (Estimated UD position) 
\EndProcedure

 \end{algorithmic}
 \end{spacing}
\end{algorithm} 

After recovering $\hat{\mathbf{x}}$ using OMP, the proposed method processes the strongest $K_{max}$ channel gains, and accepts LED ID as nearby only if it is sufficiently close, i.e., within $d_{th}$ to the running position estimate $\hat{\mathbf{u}}$. The subroutine PROX($\{\mathbf{z}_i\}_{i \in \hat{\mathcal{S}}_{\mathbf{\lambda}}}$) estimates the UD position using proximity method, i.e., as an average point of the set of LED positions $\{\mathbf{z}_i\}_{i \in \hat{\mathcal{S}}_{\mathbf{\lambda}}}$.

\section{Results and discussion}
\label{sec:4}

In this section, we present numerical experiments for an open-plan office space with square-shaped floor area of side length $50$m and height $h=3$m. LED sources are fixed on an equidistant $n_{LED} \times n_{LED}$ grid on a ceiling plane, providing for $N=n_{LED}^2$ LEDs. LED beams are shaped to provide circular coverage areas $\mathcal{A}_i$ of radius $r$. Each LED transmits a random binary LED ID signature of length $M$, thus $\mathbf{S}$ is an $M \times N$ matrix of i.i.d. equally-likely $\{0,1\}$ entries. LED and PD parameters are: $\Phi_{1/2}$ = 30$^{\circ}$, ${T_{s}}(\psi)$~ = 1, $n_{r}$ = 1.5, $\psi_c$ = 80$^{\circ}$ and $A_{r}$ = 1 $\mathrm{cm^2}$, providing the channel gains based on the model described in \eqref{eq2}. Optical power $P_{LED}$ is adjusted to provide average horizontal illuminance of $I_h=560$ lx. The noise power is obtained from the fixed and pre-selected received signal-to-noise ratio (SNR).

We first evaluate the case with the fixed geometry: i) fixed number of LEDs: $n_{LED}=25$ ($N=625$), and ii) fixed radius $r=4$m of $\mathcal{A}_i$. For each simulation run, we randomly distribute $N_{u}=1000$ UDs on the floor plane. For each UD, we generate its received signal $\mathbf{y}$ corrupted by an additive Gaussian noise of a given SNR. The UD applies Algorithm \ref{VLC-Pos} to estimate the set of nearby LEDs $\hat{\mathcal{S}}_{\mathbf{\lambda}}$ and its own position $\hat{\mathbf{u}}$. From the obtained estimates and known true values, we calculate the Euclidean-distance position error between position estimate $\hat{\mathbf{u}}$ and true position $\mathbf{u}$, and support recovery error (SRE) defined as the cardinality of symmetric difference between true and estimated support set $\hat{\mathcal{S}}_{\mathbf{\lambda}}$. The results are averaged over $N_u$ random UD instances. 

\begin{figure}[htbp]
\vspace{-10pt}
\center
  \includegraphics[width=0.5\textwidth]{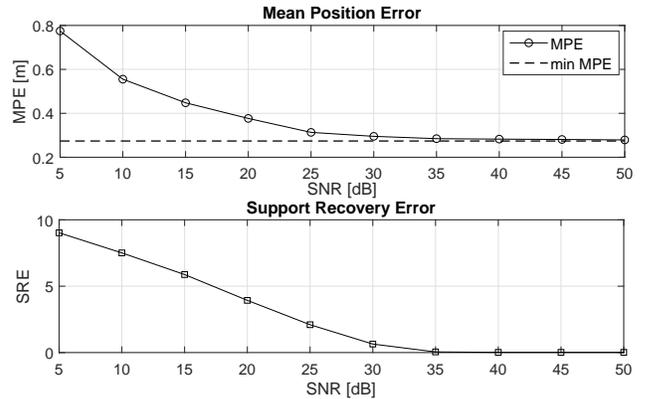}
\caption{MPE and SRE of position estimation for fixed $M=200$ as a function of $SNR$.}
\label{Fig:2} 
\vspace{-10pt}
\end{figure} 
 
In the first set of experiments (Fig. \ref{Fig:2}), we fixed the length of LED ID signatures to $M=200$ bits, and evaluated the Mean (Euclidean distance) Position Error (MPE) and SRE as a function of SNR. We note that the MPE of CS-based positioning gradually approaches the lower bound (min MPE), i.e., the precision of proximity method assuming that the set of LEDs covering the UD is correctly recovered: $\hat{\mathcal{S}}_{\mathbf{\lambda}}=\mathcal{S}_{\mathbf{\lambda}}$. Similarly, the reconstructed $\hat{\mathcal{S}}_{\mathbf{\lambda}}$ becomes correct, i.e., SRE$ \rightarrow 0$, as SNR increases. For SNR$ > 35$dB, both MPE/SRE achieve minimal values. For a given geometry, the min MPE of proximity method $\sim 0.27$m, while sparsity $K$ is position-dependent and varies between $10$ and $14$ (note that the number of LEDs covering a given randomly placed point is a non-trivial generalization of a classical Gaussian circle problem).

\begin{figure}[htbp]
\vspace{-12pt}
\center
  \includegraphics[width=0.5\textwidth]{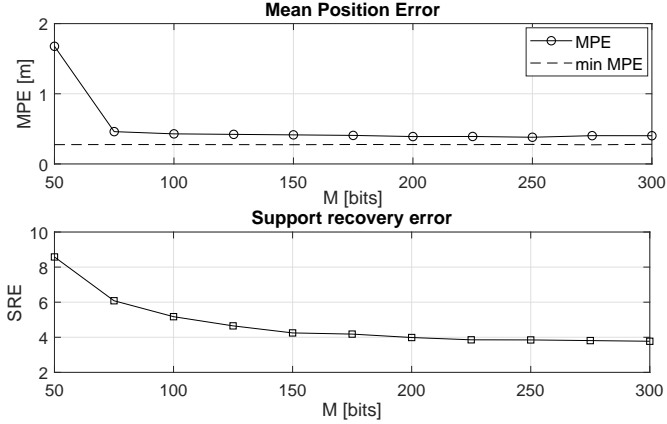}
\caption{MPE and SRE of position estimation for fixed SNR$=20$ dB as a function of $M$.}
\label{Fig:3} 
\vspace{-7pt}
\end{figure}

In the next set of experiments (Fig. \ref{Fig:3}), we fixed SNR$=20$ dB and varied LED ID signature length $M$. We note that $M$ can be reduced down to $M=75$ without almost no loss in MPE and slight loss in SRE. Note also that, at SNR$=20$ dB, MPE cannot go below MPE $\sim 0.4$m, thus further decrease towards MPE$_{min}$ is possible only by increasing SNR.  

\begin{figure}[htbp]
\vspace{-12pt}
\center
  \includegraphics[width=0.5\textwidth]{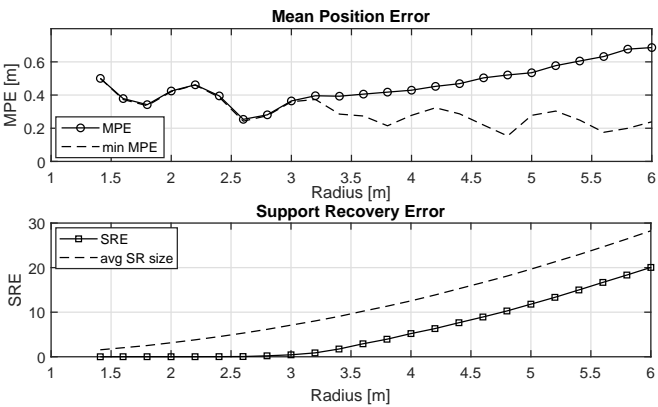}
\caption{MPE and SRE of position estimation for fixed $M=100$ and SNR$=20$ dB as a function of radius $r$.}
\label{Fig:4} 
\vspace{-10pt}
\end{figure}

We continue experiments by fixing $M=100$ and SNR$=20$dB, but changing the system geometry by varying the LED coverage radius $r$. Fig. \ref{Fig:4} reveals interesting non-monotone behavior of min MPE of proximity method as a function of $r$. Our positioning method is able to match the min MPE up to $r \sim 3.2$m for a given set of parameters. Extending this behavior to larger values of $r$ would require increase of SNR.

\begin{figure}[htbp]
\center
  \includegraphics[width=0.5\textwidth]{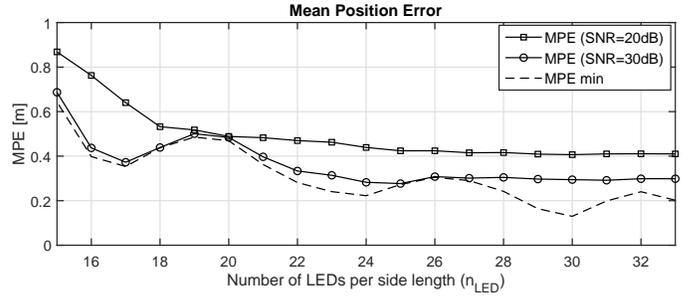}
\caption{MPE of position estimation for fixed $M=100$ and SNR$=\{20,30\}$ dB as a function of radius $n_{LED}$.}
\label{Fig:5} 
\vspace{-10pt}
\end{figure}

Finally, in Fig. \ref{Fig:5}, for a fixed $M=100$, $r=4$m, and two values of SNR$=\{20,30\}$dB, we change the geometry via changing the LED density $n_{LED}.$ Due to space constraints, we present only MPE results, and note the similar behaviour as in the previous cases, where increased SNR shifts the performance towards min MPE bound.  

\section{Conclusion}

In this work, we presented a VLC-based positioning approach using compressed sensing for signal separation and proximity method for position estimation. The presented solution is of low-complexity while performing close to the bounds achievable using proximity method for a range of parameters.

\end{document}